\relax
\documentclass[letterpaper]{article} 
\usepackage{aaai22}  
\usepackage{times}  
\usepackage{helvet}  
\usepackage{courier}  
\usepackage[hyphens]{url}  
\usepackage{graphicx} 
\usepackage{multirow}
\usepackage{stfloats}
\usepackage{titling}
\usepackage{authblk}
\usepackage{natbib}  
\usepackage{caption} 
\DeclareCaptionStyle{ruled}{labelfont=normalfont,labelsep=colon,strut=off} 
\usepackage{algorithm}
\usepackage{algorithmic}

\usepackage{amsmath}
\usepackage{mathrsfs}
\usepackage{bm}

\usepackage{newfloat}
\usepackage{listings}
\floatstyle{ruled}
\newfloat{listing}{tb}{lst}{}
\floatname{listing}{Listing}
\title{Hide and Seek (HaS): A Lightweight Framework for Prompt Privacy Protection}
\author{
    Yu Chen,
    Tingxin Li,
    Huiming Liu,
    Yang Yu \thanks{The corresponding author.},
}
\affil{
    Xuanwu Lab, Tencent \\
    alohachen, tingxinli, huimingliu, tkyu@tencent.com
}

\usepackage{bibentry}

\begin{document}

\maketitle

\begin{abstract}
Numerous companies have started offering services based on large language models (LLM), such as ChatGPT, which inevitably raises privacy concerns as users' prompts are exposed to the model provider. Previous research on secure reasoning using multi-party computation (MPC) has proven to be impractical for LLM applications due to its time-consuming and communication-intensive nature. While lightweight anonymization techniques can protect private information in prompts through substitution or masking, they fail to recover sensitive data replaced in the LLM-generated results.

In this paper, we expand the application scenarios of anonymization techniques by training a small local model to de-anonymize the LLM's returned results with minimal computational overhead. We introduce the HaS framework, where ``H(ide)" and ``S(eek)" represent its two core processes: hiding private entities for anonymization and seeking private entities for de-anonymization, respectively. 

To quantitatively assess HaS's privacy protection performance, we propose both black-box and white-box adversarial models. Furthermore, we conduct experiments to evaluate HaS's usability in translation and classification tasks. The experimental findings demonstrate that the HaS framework achieves an optimal balance between privacy protection and utility.

\end{abstract}

\section{1 \quad Introduction}
Large language models (LLMs) have gained significant attention in recent years due to their remarkable performance in various natural language processing tasks, such as summarization, translation, and question-answering. As a result, an increasing number of companies have started offering services based on LLMs, such as ChatGPT. However, the widespread adoption of LLMs has raised privacy concerns, as users' prompts are exposed to the model provider. In many cases, these prompts may contain sensitive information that users would prefer not to disclose.

Previous research on secure reasoning using multi-party computation (MPC) has attempted to address these privacy concerns. However, MPC-based approaches are impractical for LLM applications due to their time-consuming and communication-intensive nature. Lightweight anonymization techniques, which protect private information in prompts through substitution or masking, provide a more efficient alternative. Unfortunately, these techniques often fail to recover sensitive data replaced in the LLM-generated results, thereby compromising the utility of the output.

To address this challenge, we propose a novel framework called Hide and Seek (HaS) for prompt privacy protection in LLMs. The HaS framework comprises two core techniques: hiding private entities for anonymization and seeking private entities for de-anonymization. More specifically, HaS achieves privacy protection by hiding private entities in the prompt through the use of anonymization modules. We experimented with two anonymization schemes: generative and label-based. The generative scheme employed a Bloomz model trained on data annotated by GPT-4, while the label-based scheme utilized existing Named Entity Recognition (NER) models. Upon receiving the results from the LLM, HaS employs a de-anonymization process through Bloomz model which trained on data annotated by GPT3.5-turbo. This approach allows for the preservation of sensitive information privacy while ensuring the utility of the generated output with minimal computational overhead. In order to quantitatively assess the privacy protection performance of HaS, we introduce both black-box and white-box attack scenarios. For these two attack scenarios, we have trained two adversary models separately to carry out decryption attacks, and then scored the privacy protection capabilities based on the attack results. Additionally, we conduct usability experiments to evaluate the performance changes in executing translation and classification tasks before and after using HaS. The results of these two sets of experiments demonstrate that HaS provides a comprehensive framework for the privacy-utility trade-off.


The main contributions of this paper are as follows:

\begin{itemize}
\item We propose a novel LLM prompt privacy protection framework HaS, which can perform anonymization and de-anonymization on privacy entities locally.

\item We propose two hide-seek schemes to protect privacy entities, and put forth two adversary models to evaluate their resilience against deciphering attacks.

\item We open-source the first privacy entity anonymization dataset and privacy entity de-anonymization dataset, as well as the anonymization and de-anonymization models\footnote{https://github.com/alohachen/Hide-and-Seek}. 
\end{itemize}

In summary, the HaS framework offers a practical and effective solution for safeguarding user privacy in LLM applications, alleviating the escalating concerns associated with the deployment of these powerful models in everyday services to a certain extent. The rest of the paper is organized as follows: Section 2 provides a literature review of existing privacy protection methods and their limitations; Section 3 presents the details of the HaS framework and its implementation; Section 4 discusses the experimental setup and evaluation results; and Section 5 concludes the paper and outlines future research directions.

\section{2 \quad Related Work}

In this section, we discuss related work in the area of privacy-preserving techniques, focusing on their applicability in protecting end-side inference privacy, supported data types, computational overhead, and deployment difficulty. We propose HaS, which provides privacy protection for free-text data, has low computational overhead, and supports local plug-and-play deployment, making it transparent to cloud-based LLM.

Existing privacy-preserving techniques include Differential Privacy (DP) \cite{dwork2006calibrating}, Multi-Party Computation (MPC) \cite{yao1982protocols}, Federated Learning (FL) \cite{mcmahan2016communication}, Data Anonymization \cite{samarati1998protecting}, and Homomorphic Encryption (HE) \cite{gentry2009fully}.

Differential Privacy (DP) primarily aims to protect individual data privacy within a dataset by adding a certain degree of random noise. It supports end-side inference privacy and can extract useful statistical information while preserving privacy. However, DP is mainly applicable to numerical data and has difficulty in handling free-text data due to its high dimensionality and complexity. The computational overhead of DP is relatively high, and its deployment requires modifications to the LLM.

Multi-Party Computation (MPC) focuses on protecting private data during collaborative computation among multiple parties, including tasks such as inference, data analysis, and data mining. MPC employs cryptographic techniques to encrypt data, ensuring that the original data is not exposed during computation. Although MPC can theoretically support free-text data, it faces challenges in terms of efficiency and scalability due to the complexity of text data. MPC has a high computational overhead and requires modifications to the LLM for deployment.

Federated Learning (FL) is a privacy-preserving technique for distributed machine learning among multiple parties. In FL, each party trains a machine learning model on their local device and then aggregates the updated model parameters on a central server. FL does not support end-side inference privacy as only model parameters are shared among parties. However, it can handle various data types, including free-text data, by mapping strings to word vectors or encodings. FL has a relatively high computational overhead and requires modifications to the LLM for deployment.

Data Anonymization] is a technique that removes or replaces sensitive information in data while preserving its structure and partial information. It supports end-side inference privacy and can handle various data types, including free-text data. Data anonymization has low computational overhead and deployment difficulty. However, it lacks the ability to restore anonymized entities to their original forms.

Homomorphic Encryption (HE) is a cryptographic technique that allows direct computation on encrypted data without the need for decryption. HE supports end-side inference privacy and can handle various data types, including free-text data. However, the implementation overhead of HE is relatively high, and its deployment requires modifications to the LLM.

In summary, our proposed HaS technique offers several advantages over existing privacy-preserving techniques, such as support for free-text data, low computational overhead, and ease of deployment. It provides a promising solution for protecting end-side inference privacy in various applications.

\section{3 \quad Our Approach}
\begin{figure}
  \centering
\includegraphics[width=0.5\textwidth]{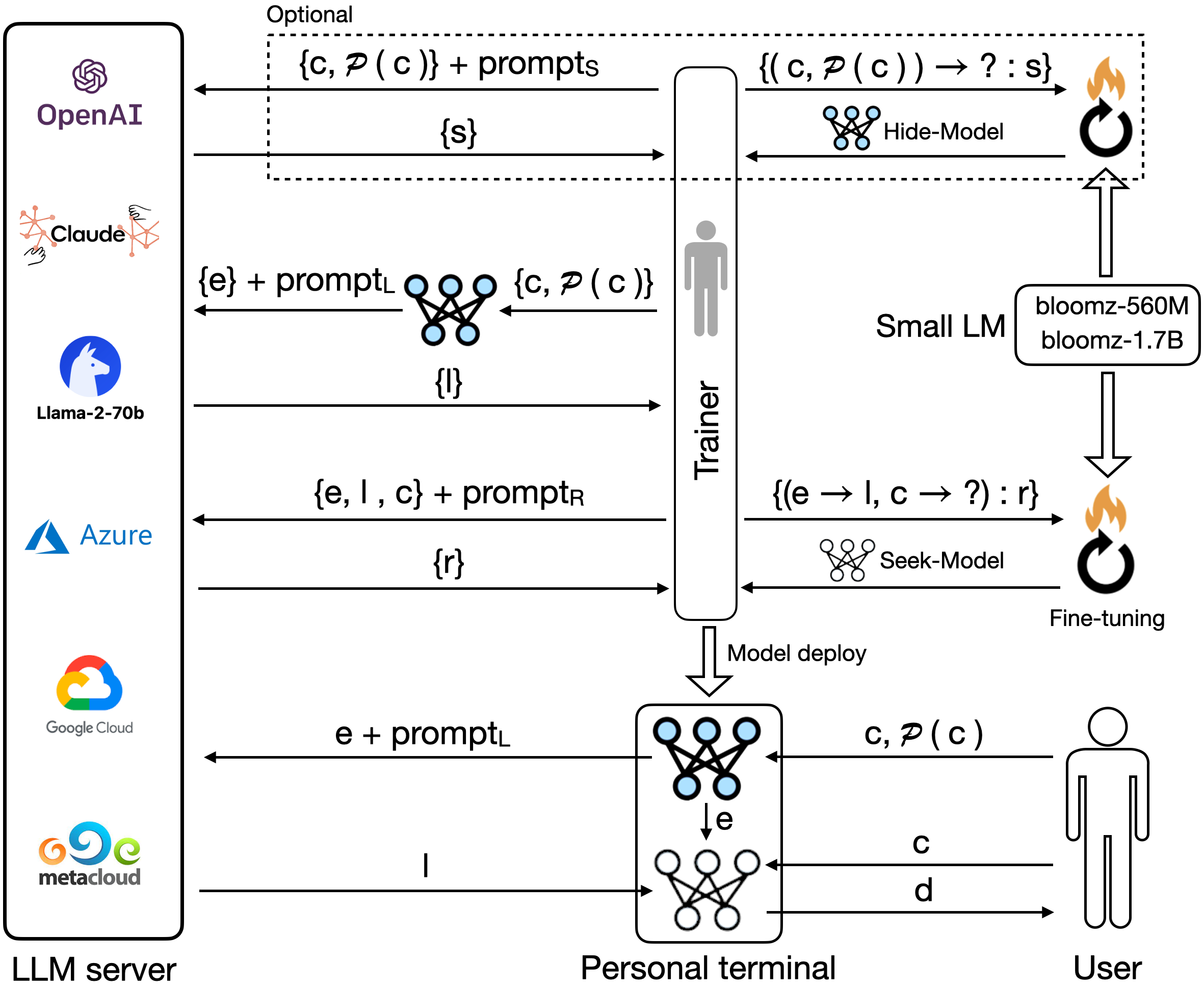}
  \caption{Overview of HaS}
  \label{fig:overview}
\end{figure}

As shown in Figure \ref{fig:overview}, HaS is divided into two phases: training and inference. The main task of the training phase is to generate training dataset by using LLM through prompt engineering. The generated training data is then used to train two small LLMs for anonymization and de-anonymization respectively. It should be noted that if the Label-Based Hide-Model scheme is adopted, there is no need to train Hide-Model, and the existing NER model can be used directly. 

The main task of the inference phase is to hide the private entities of the user-submitted text with the Hide-Model deployed on the personal terminal (e.g., cell phone, tablet, PC, etc.); output the de-anonymized text after hiding the private entities; and then send the de-anonymized text to the LLM in the cloud and to the Seek-Model, which is also deployed in the local area, respectively. The input of Seek-Model is the text returned by LLM, the anonymized text from Hide-Model and the original input, and the output is the de-anonymized result.

\subsection{3.1 Notation.}
First we introduce the major roles involved in the HaS framework:

\noindent \textbf{User}: Ordinary users who use LLM services for text conversion tasks. The text conversion tasks include but are not limited to machine translation, polishing, summarization, and other tasks, but do not include knowledge retrieval, article continuation, and the like.

\noindent \textbf{LLM server}: LLM services provided by AI and cloud vendors. Users access LLM services in the cloud through APIs or web interfaces such as ChatGPT, GPT3.5-turbo API, GPT4 API, Claude, PaLM, etc.

\noindent \textbf{Personal terminal}: Mobile phones, tablets, personal computers and other terminal devices. These edge computing devices have a certain amount of arithmetic power to run small parametric quantities of models, such as 560M or 1.7B sized models.

\noindent \textbf{Trainer}: HaS builder, responsible for training Hide-Model and Seek-Model on its own machine and then deploying them on the user's Personal terminal.

\noindent The symbols involved in Figure \ref{fig:overview} are presented next.

\noindent \(\bm{\mathcal{P}}\): Privacy entity extraction operation where the privacy entity is a named entity of the following type: 

\textit{DATE}: Absolute or relative dates or periods

\textit{MONEY}: Monetary values, including unit

\textit{PERCENT}: Percentage, including ``

\textit{QUANTITY}: Measurements, as of weight or distance

\textit{TIME}: Times smaller than a day

\textit{GPE}: Countries, cities, states

\textit{LOC}: Non-GPE locations, mountain ranges, bodies of water

\textit{PERSON}: People, including fictional

\textit{WORK\_OF\_ART}: Titles of books, songs, etc.

\textit{ORG}: Companies, agencies, institutions, etc.

\textit{NORP}: Nationalities or religious or political groups

\textit{LAW}: Named documents made into laws.

\textit{FAC}: Buildings, airports, highways, bridges, etc.

\textit{}{LANGUAGE}: Any named language

\noindent \bm{$s$}: text anonymized by LLM

\noindent \bm{$l$}: text processed by LLM

\noindent \bm{$r$}: text de-anonymized by LLM

\noindent \bm{$e$}: text anonymized by Hide-Model

\noindent \bm{$d$}: text de-anonymized by Seek-Model

\noindent Lastly, let me introduce the guiding prompt involved in HaS for LLM work:

\noindent \bm{$prompt_{S}$}:
\\

\fbox{%
  \begin{minipage}{0.9\linewidth}

Prompt for guiding LLM to perform anonymization.

Substitute given words in the text into other random words.

Text: \{input\_text\}

Given words: {private entities}

Substituted text:

\{task\_type\}:
  \end{minipage}
}

\noindent \bm{$prompt_{L}$}:
prompt for guiding LLM to process text.

\noindent \bm{$prompt_{R}$}:

prompt for guiding LLM to perform de-anonymization.
\\

\fbox{%
  \begin{minipage}{0.9\linewidth}
Input: \{obscured\_text\}

\{task\_type\}: \{obscured\_api\_response\}

Input: \{input\_text\}

\{task\_type\}:
  \end{minipage}
}

where task\_type can either be ``Translate", ``Abstract" or ``Polish".

\subsection{3.2 Pipelines.}
Since there are two types of Hide-Models: generative and label-based, they correspond to two different training and inference pipelines.

\noindent \textbf{Generative Pipeline:}

\textbf{STEP 1}: The trainer needs to prepare a domain-specific corpus. For each data entry $c$ in the corpus, use P to extract the private entity $P(c)$. Then upload  $c$ and $P(c)$ into the LLM according to the format of $prompt_{S}$, and obtain the returned results $s$. 

Input text $c$: The FBI (Federal Bureau of Investigation) is currently investigating a cyber attack on a major corporation that occurred on August 10, 2023. The breach took place in the company's headquarters located in Washington DC. The FBI suspects that the attack was carried out by a foreign government.

Private entities $P(c)$: [`FBI', `August 10, 2023', `Washington DC']

Substituted text $s$: The CIA (Central Intelligence Agency) is currently investigating a cyber attack on a major corporation that occurred on September 15, 2025. The breach occurred in the company's headquarters located in New York City. The CIA suspects that the attack was carried out by a foreign government.

\textbf{STEP 2}: 
Utilize the obtained $c$,$P(c)$ and $s$ from the previous step to train a Small LLM named Hide-Model. The training template is as follows:
\\

\fbox{%
  \begin{minipage}{0.9\linewidth}
Substitute given words in the text into other random words.

Text: $c$

Given words: $P(c)$

Substituted text: $s$
  \end{minipage}
}
\\

\textbf{STEP 3}: Use the Hide-Model obtained from the previous training step to perform batch anonymization on the corpus' $c$ and its $P(c)$, then submit the anonymized results $e$ in batches to the LLM for processing as specified by $prompt_{L}$, obtaining the LLM's return results $l$.

\textbf{STEP 4}: 

Upload $c$ as the ``input\_text", $e$ as the ``obscured\_text" and $l$ as the ``obscured\_api\_response" into the LLM according to the format of $prompt_{R}$, and obtain the returned results $r$. 

\textbf{STEP 5}: 

Utilize the obtained $e$,$l$ and $c$ from the previous step to train a Small LLM named Seek-Model. The training template is as follows:
\\

\fbox{%
  \begin{minipage}{0.9\linewidth}
Input: $e$

Translate: $l$

Input: $c$

Translate: $r$
  \end{minipage}
}
\\

\textbf{STEP 6}: 
The trainer will deploy the trained Hide-Model and Seek-Model to the user's personal terminal in the form of an app or plugin.

\textbf{STEP 7}: 
After the user submits the original text $c$ to be processed, first manually specify or automatically recognize the privacy entities $P(c)$, and then use the deployed Hide-Model to perform anonymization. The anonymized text $e$ is sent to the LLM for processing according to the instructions of $prompt_{L}$. Finally, the Seek-Model outputs the de-anonymized result {d} based on {l}, {c}, and {e}.

\noindent \textbf{Label-Based Pipeline:}
Most of the steps in the Label-Based Pipeline are the same as those in the Generative Pipeline. The main difference is that the Label-Based Pipeline does not require training a Hide-Model in STEP1 and STEP2, but can instead use an existing NER Model as the Hide-Model directly. Therefore, the Label-Based Pipeline starts from STEP3 of the Generative Pipeline, and an example of the new STEP3 is as follows:

Input text $c$: The FBI (Federal Bureau of Investigation) is currently investigating a cyber attack on a major corporation that occurred on August 10, 2023. The breach took place in the company's headquarters located in Washington, D.C. The FBI suspects that the attack was carried out by a foreign government.

Private entities $P(c)$: [`FBI', `August 10, 2023', `Washington, D.C.']


Anonymized text $e$: The \textless ORG\textgreater{} (\textless ORG\textgreater{}) is currently investigating a cyber attack on a major corporation that occurred on \textless DATE\textgreater{}. The breach took place in the company's headquarters located in \textless GPE\textgreater{}, \textless GPE\textgreater{} The \textless ORG\textgreater{} suspects that the attack was carried out by a foreign government.



\subsection{3.3 \quad Adversary model}
\begin{figure}
  \centering
\includegraphics[width=0.5\textwidth]{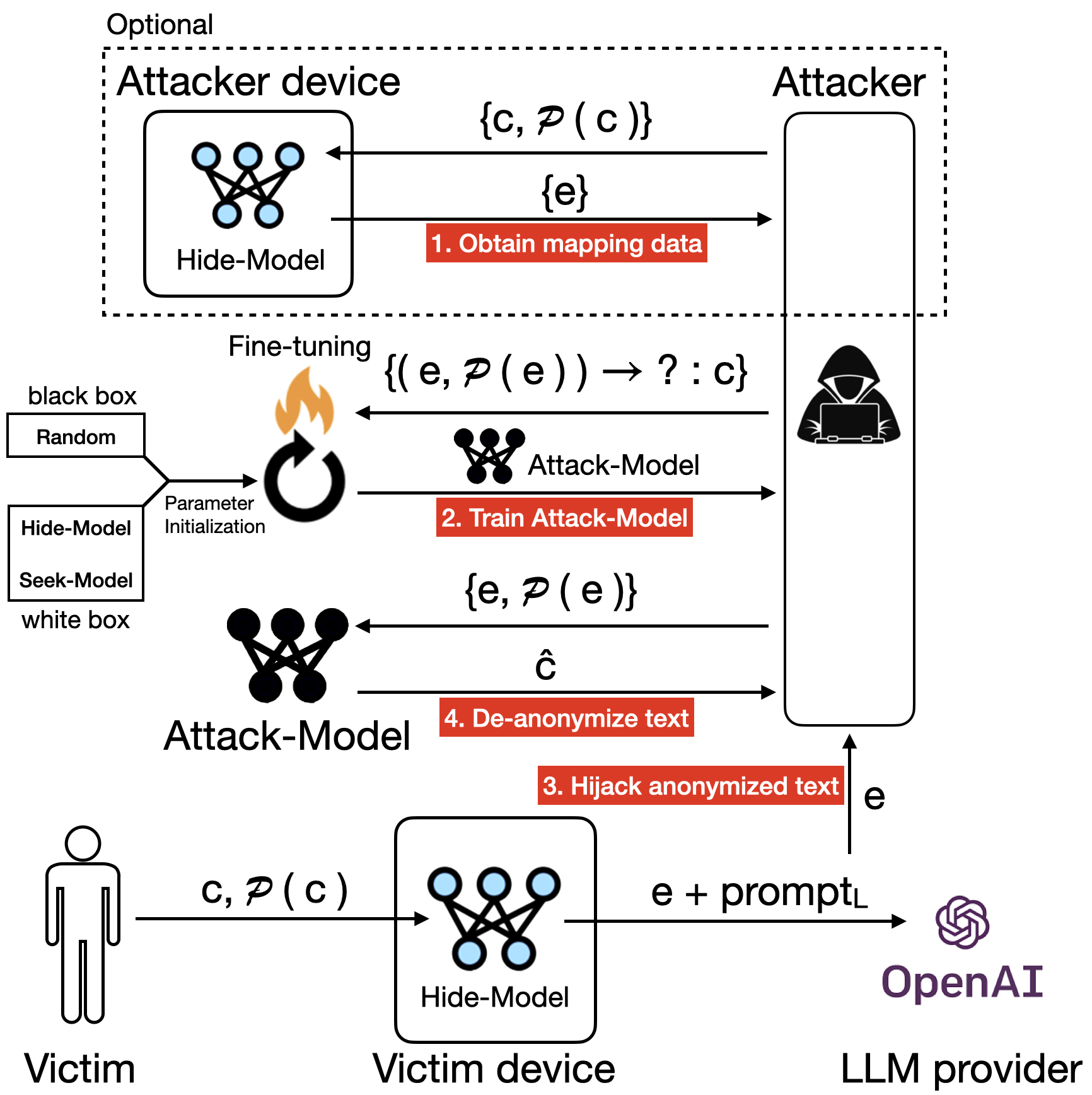}
  \caption{Adversary model}
  \label{fig:attack}
\end{figure}

In order to demonstrate the security of HaS when encountering active attacks from adversaries, we propose a realistic adversary model. In this adversary model:

STEP1. Attackers can steal the anonymous text uploaded by the victim to the LLM provider. For example, both the LLM provider itself and attackers who can hijack the network communication channel meet this condition.

STEP2. Optionally, the attacker can obtain the Hide-Model deployed on the victim's device, submit a large number of anonymous requests to it, and record these requests $c$,$P(c)$, and their corresponding anonymous results $e$. This step is optional because, for the Label-based Hide Model, the results of NER can be used directly for label type annotation.

STEP3. The attacker uses the $c$, $P(c)$, and $e$ obtained from the above steps (or directly from NER) to train the attack model. The initial weights of the attack model can be randomly initialized or use the Hide-model/Seek-model, with the former being called black-box attack and the latter being called white-box attack. In the white-box, attacker initialize the parameters using either the Hide-Model or Seek-Model to investigate whether the prior knowledge stored in the model parameters is beneficial for the attack. The sets of prompt words for training the attack model are as follows:
\\

\noindent Black-box attack towards label-based Hide-Model: 
\\
\fbox{%
  \begin{minipage}{0.9\linewidth}

Recover the following text with label masks.

Text: $e$

Recovered text: $c$

  \end{minipage}
}
\\

\noindent Black-box attack towards generative Hide-Model: 
\\
\fbox{%
  \begin{minipage}{0.9\linewidth}

Recover given words in the text into original words.

Text: $e$

Given words: $P(e)$

Recovered text: $c$

  \end{minipage}
}
\\

\noindent White-box attack towards generative/label-based Hide-Model that initialized by the parameters of victim's Seek-Model: 
\\
\fbox{%
  \begin{minipage}{0.9\linewidth}
Input: $e$

Translation: the translation of $e$

Input: $c$

  \end{minipage}
}
\\
Our main consideration in designing the attack prompt is to make them as similar as possible to the prompt used in training Hide-Model and Seek-Model, which we thought might maximise the use of information from the existing weights.

STEP4: Finally, attackers use the trained attack model to perform a de-anonymisation attack on the intercepted anonymous prompt words.

\section{4 \quad Experiments}

\subsection{4.1 \quad Experimental Setup}

In this section, we are going to introduce the detailed settings of our experiments. 

For the label-based hider-model, we adopt the NER pipeline of SpaCy \footnote{https://spacy.io}. For the base model of generative hider-model, seeker-model, and adversarial model, we use the version of BLOOMZ \cite{muennighoff2022crosslingual} implemented by huggingface transformer\footnote{https://github.com/huggingface/transformers}. We fine-tune the model on our sythetic dataset using QLoRA introduced by \cite{dettmers2023qlora}, with huggingface PEFT\footnote{https://github.com/huggingface/peft} for parameter tuning and bitsandbytes\footnote{https://github.com/TimDettmers/bitsandbytes} for model quantization. 

During training, we use AdamW \cite{loshchilov2018decoupled} optimizer and set the learning rate as $2 \times 10^{-4}$. We also add a linear learning rate scheduler with 5\% of the training steps as warm up stage. For the generative hider-model and adversarial-model, we set max length as 1250 and batch size as 24. For the seeker-model, we set max length as 1550 and batch size as 16. During inference, we use beam search with beam size 3 to select optimal token sequence. Furthermore, we add repetition penalty with $\theta=3$ \cite{keskar2019ctrl} because we discovered that such penalty makes the model replace entities more proactively.

\subsection{4.2 \quad Datasets}
As introduced earlier, we use synthetic data to train the generative hider model, seeker model, and adversarial model. The base corpus is a news summarization corpus\footnote{https://www.kaggle.com/datasets/sbhatti/news-summarization}. We select 19703 samples from the corpus as raw input for hider model and seeker model, and generate training data with OpenAI GPT-3.5 or GPT-4 API\footnote{https://openai.com/blog/openai-api}. 80\% of the data is used for training and 20\% is used for model testing. To simulate attacking, we select another 30000 samples from the news summarization corpus to train the adversarial model, and test attacking on the same test set. Specially, for the evaluation on classification task, we use BBC news classification dataset\footnote{https://www.kaggle.com/competitions/learn-ai-bbc}, which contains 1490 labeled samples so that we have ground truth to do evaluation. 

\subsection{4.3 \quad Results}
In this section, we are going to demonstrate and analyze our experiment results.

Firstly, we evaluate the effect of privacy protection using adversarial model. To make the comparison fair, we use adversarial model of size 1.7 billion for all of the experiments. We measure the effect of attacking by calculating $1 - s(c, \hat{c})$, where $s(c, \hat{c})$ is the similarity between the original text $c$ and recovered output $\hat{c}$. The similarity is obtained from the matching function of Python package difflib\footnote{https://docs.python.org/3/library/difflib.html}. The metric is designed to increase when the attacking model find it hard to recover the original text. The results are shown in table \ref{tab:table1}. Overall, the label-based method gives better protection compared to the generative method. For the generative method, larger model gives better protection. The results also show that, even the hackers got our model weights, it will not help them to break our model.

\begin{table*}[tp]
    \centering
        \begin{tabular}{c c c c}
          \hline
          \textbf{Strategy} & \textbf{Black} & \textbf{White (hider)} & \textbf{White (seeker)}\\
          \hline
          Generative (560m) & 0.3121 & 0.3147 & 0.3212 \\

          Generative (1b7) & 0.3259 & \textbf{0.3230} & 0.3401 \\

          Label-based & \textbf{0.4070} & \textbackslash & \textbf{0.3999} \\
          \hline
        \end{tabular}
    \caption{Test results of privacy protection effect}
    \label{tab:table1}
\end{table*}

\begin{table*}[tp]
    \centering
        \begin{tabular}{c c c c | c c c | c}
          \hline
          \textbf{Strategy} & \textbf{P (macro)} & \textbf{R (macro)} & \textbf{F1 (macro)} & \textbf{P (micro)} & \textbf{R (micro)} & \textbf{F1 (micro)} & \textbf{$\Delta$ F1 (micro)}\\
          \hline
          No obscure & 91.76 & 91.36 & 91.09 & 92.18 & 91.14 & 91.20 & \textbackslash \\
          \hline
          Generative (560m) & \textbf{91.63} & \textbf{91.60} & \textbf{91.20} & \textbf{92.20} & \textbf{91.34} & \textbf{91.36} & \textbf{+0.17\%} \\

          Generative (1b7) & 91.45 & 91.25 & 90.86 & 92.00 & 90.94 & 90.98 & -0.24\% \\

          Label-based & 91.11 & 90.46 & 90.07 & 91.64 & 90.07 & 90.14 & -1.16\% \\
          \hline
        \end{tabular}
    \caption{Test results of privacy budget on classification task}
    \label{tab:table2}
\end{table*}

\begin{table*}[!htbp]
    \centering
        \begin{tabular}{c c c c c c c c | c}
          \hline
          \textbf{Strategy} & \textbf{Setting} & \textbf{ROUGE-1} & \textbf{ROUGE-2} & \textbf{ROUGE-L} & \textbf{BLEU-2} & \textbf{BLEU-4} & \textbf{METEOR} & \textbf{$\Delta$ METEOR}\\
          \hline
          No obscure & \textbackslash & 64.18 & 39.08 & 54.69 & 51.81 & 38.55 & 61.95 & \textbackslash \\
          \hline
          Generative & \multirow{2}{*}{560m} & 60.73 & 36.07 & 51.73 & 49.05 & 35.50 & 58.45 & -5.65\% \\

          Label-based & ~ & 60.58 & 34.80 & 50.89 & 49.80 & 36.05 & 58.94 & -4.86\% \\
          \hline
          Generative & \multirow{2}{*}{1b7} & \textbf{62.36} & \textbf{37.37} & \textbf{53.33} & 50.80 & 37.16 & 60.52 & -2.32\% \\
            
          Label-based & ~ & 62.16 & 36.42 & 52.70 & \textbf{51.83} & \textbf{37.95} & \textbf{61.43} & \textbf{-0.85\%} \\
          \hline
        \end{tabular}
    \caption{Test results of privacy budget on translation task}
    \label{tab:table3}
\end{table*}

Then, we evaluate the privacy budget of different methods through a classification task. The results of classification task are shown as follows. As shown in table \ref{tab:table2}, the results indicate that the generative method keeps the semantics of the text better than label-based method and cost less privacy budget.

\begin{figure}
  \centering
  \includegraphics[scale=0.55]{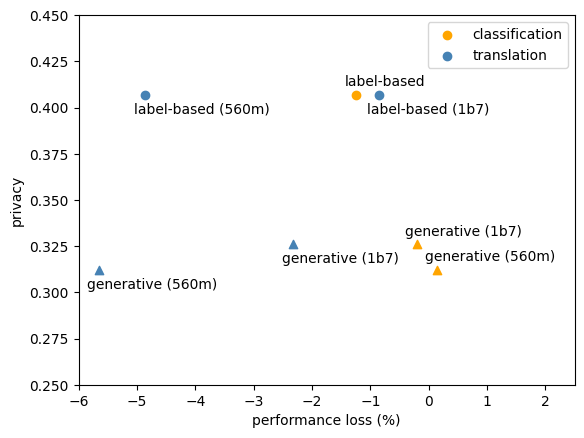}
  \caption{Study of privacy protect \& privacy budget}
  \label{fig:compare}
\end{figure}

Afterwards, we evaluate the performance of seeker model using translation task. We apply seeker model to revert obscured translation into the translation of original text. To evaluate the privacy budget, we use DeepL\footnote{https://www.deepl.com} as reference translation. The results are shown in table \ref{tab:table3}. After applying seeker model, the privacy budget of label-based hider model become lower than generative hider model. This indicates that the seeker model performs better on label-based obscured text than the synthetic text so it can revert the translation more accurately even if label-based obscuring tends to introduce more information loss.

In addition, it is notable that, if we study the relation between privacy protection effect and privacy budget, we can intuitively discover that better protection costs more privacy budget for the classification, as shown in figure \ref{fig:compare}. However, for the translation task, the regulation is different because the seek model performs better on label-based obscured text.

\section{5 \quad Discussion}
\subsection{Limitations.}

\begin{itemize}
    \item Although HaS can anonymize privacy entity words, it cannot hide other words and sentence structures, such as verbs and other non-named entity words, and therefore cannot provide full privacy protection capabilities.
    \item Since HaS replaces the entity words in the prompt, it leads to tasks that rely on the precise semantics of the replaced entity words, such as knowledge retrieval and text continuation, being unable to use HaS.
\end{itemize}

\subsection{Future Work.}
\begin{itemize}
    \item Use RLHF or RLAIF to adjust the training of the generative hiding scheme, in order to reduce the probability of synonyms being output and improve the model's resistance to decryption attacks.
    \item Remove the NER step and directly generate the anonymized text end-to-end, thereby further reducing the storage and operational overhead of HaS.
    \item Increase the amount of training data to directly fine-tune the model without using Lora, and support more languages.
\end{itemize}

\section{6 \quad Conclusion}
In this paper, we have presented the HaS framework, a novel approach for protecting privacy in large language model applications. By incorporating anonymization and de-anonymization techniques, HaS effectively balances privacy protection and utility. Our proposed hide-seek schemes demonstrate resilience against deciphering attacks, as evaluated by our black-box and white-box adversarial models. Additionally, we have open-sourced privacy entity anonymization and de-anonymization datasets, along with the corresponding models, to encourage further research in this area.

Despite its promising results, the HaS framework has certain limitations, such as its inability to provide full privacy protection for non-named entity words and its inapplicability to tasks that rely on precise semantics of replaced entity words. Future work will focus on addressing these limitations by improving the generative hiding scheme, exploring end-to-end anonymization techniques, and expanding the framework to support more languages.

By advancing the field of privacy-preserving LLM applications, the HaS framework paves the way for more secure and private interactions with large language models, ensuring that users can benefit from these powerful tools without compromising their privacy.

\bibliography{aaai22}

\appendix

\end{document}